\documentclass[12pt]{article}
\usepackage[margin=1.0in]{geometry}
\usepackage[utf8x]{inputenc}

\usepackage{apacite}
\usepackage{graphicx}
\usepackage{amssymb}
\usepackage{natbib}
\usepackage{epstopdf}

\usepackage[normalem]{ulem}
\usepackage{color}

\title{\normalsize \textbf{A Beginner's Guide to Line Intensity Mapping Power Spectra}}
\author{Trevor M. Oxholm\footnote{TMO would like thank Dr. Eric Switzer and Prof. Peter Timbie for useful comments on the draft and for contributions to his understanding of LIM. He also gratefully acknowleges support from the Wisconsin Space Grant Consortium Graduate \& Professional Research Fellowship for generous support enabling this research, as well as other projects referenced in this proceeding.}}
\date{Department of Physics, University of Wisconsin-Madison, Madison, WI \\
October 2021}

\begin{document}

\pagestyle{empty}

\clearpage\maketitle
\thispagestyle{empty}

\begin{abstract}
    \footnotesize Line intensity mapping (LIM) is an emerging technique in measuring galaxy evolution and the large-scale structure of the universe. LIM surveys measure the cumulative emission from all galaxies emitting a given line at a particular redshift, which trace the distribution of dark matter throughout the universe. In this proceeding, we provide an introduction to LIM modeling, focusing on power spectrum calculation. Beyond these calculations, we describe how these power spectra may be used to constrain properties of galaxy evolution and large-scale structure cosmology. Throughout, we use the anticipated EXCLAIM signal of ionized carbon ([CII]) at redshift $z=3$ as a case study. Our goal is to provide a starting point to non-experts, e.g. upper-level undergraduate and graduate students familiar with the basics of cosmology, with the tools necessary to understand the literature and generate LIM power spectrum models themselves, while also describing a wide array of literature for continued studies. 
\end{abstract}


\textbf{Introduction}

\begin{figure}
    \centering
    \includegraphics[width=\textwidth]{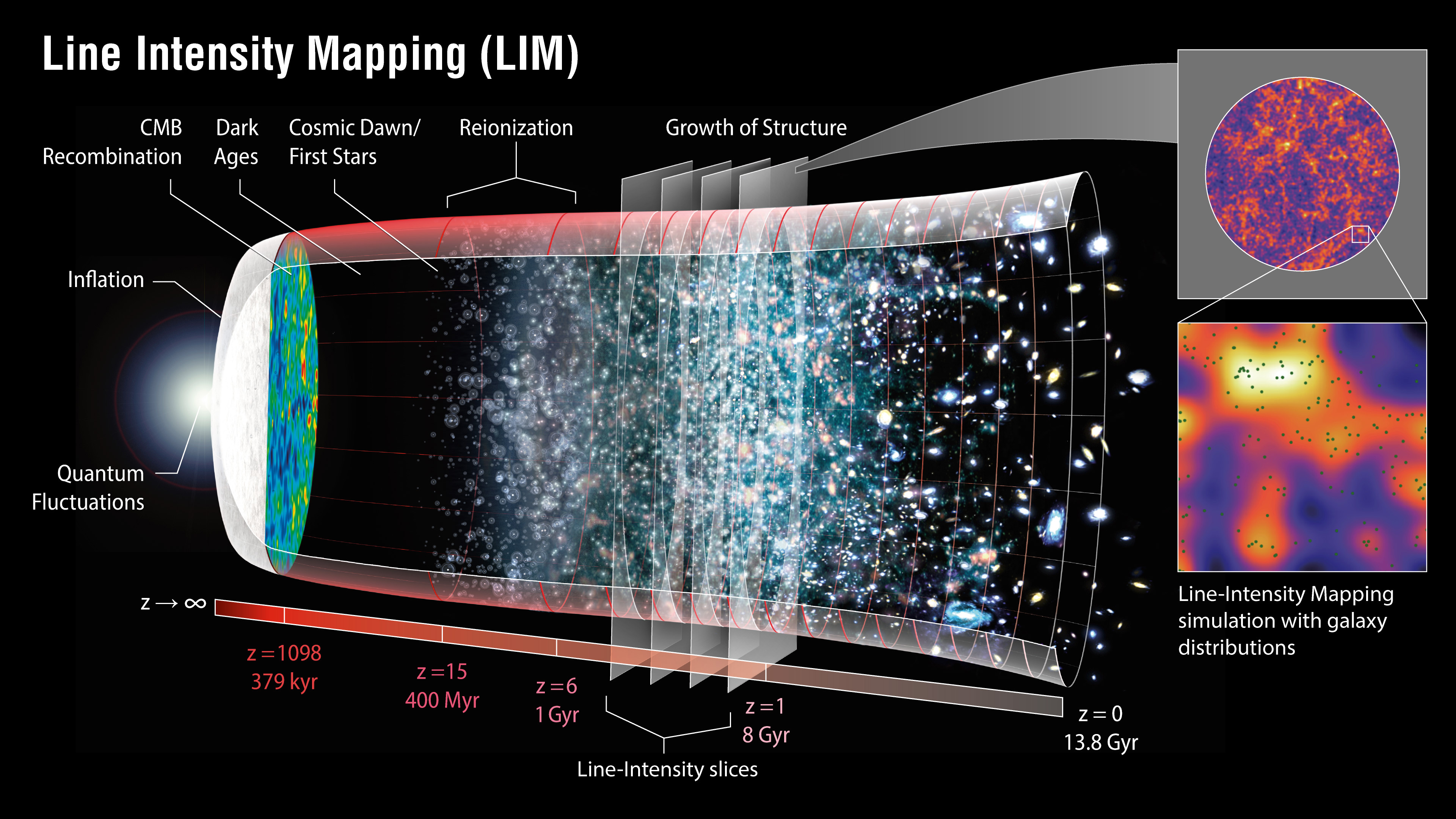}
    \caption{Overview of the line intensity mapping technique. Plot obtained from the NASA-LAMBDA archive.}
    \label{fig:LIM_overview}
\end{figure}

Line intensity mapping is an emerging technique in large-scale structure cosmology and the physics of galaxy evolution. A line intensity map traces the cumulative emission of a given emission line from all galaxies in a target region. The target galaxies trace the matter overdensity, allowing for an inference of the large-scale structure of dark matter. The target intensity (or rather, its overdensity) is therefore a biased tracer of the matter overdensity, with a multiplicative bias proportional to the mean intensity of the target line. Furthermore, by comparing the observed frequency $\nu_{\rm obs}$ to the rest-frame frequency $\nu_{\rm rest}$ of the target line, we are able to obtain precise redshift information through the relation $z = \nu_{\rm rest}/\nu_{\rm obs}-1$. This allows for unambiguous spectroscopy, while also determining the mean intensity from all galaxies emitting at the observed redshift, as demonstrated in Figure  \ref{fig:LIM_overview}. By providing a three-dimensional tomographic datacube, line intensity maps are a treasure trove for cosmological and astrophysical studies.

From a cosmological standpoint, intensity mapping surveys may access much of the same information as traditional galaxy redshift surveys such as SDSS \citep{york2000}, DES \citep{abbott2016}, BOSS \citep{dawson2012}, etc. This information includes structure growth through the measurement of large-scale redshift-space distortions. Intensity mapping surveys may also be used to measure baryon acoustic oscillations \citep{chang2008}, allowing for inference of the dark energy equation of state and cosmological parameters. Furthermore, they may also be used to measure signatures beyond the standard cosmological model, including primordial non-Gaussianity \citep{liu2021} and signatures of dark matter annihilation \citep{bernal2021}.


Intensity mapping surveys may also be used to unlock mysteries of galaxy formation and evolution, as well as the cosmic star formation history. Target emission lines, such as neutral hydrogen (HI), singly-ionized carbon ([CII]), and carbon monoxide lines (CO) trace different properties of galaxies. HI and [CII] trace the signatures of star formation, and the latter is an effective proxy for the total infrared galaxy luminosities. Furthermore, the lower transition levels of CO trace cold gas, the fuel for star formation \citep{bolatto2013}. Numerous other lines including Lyman-alpha, [OIII], H$\alpha$, and H$\beta$ may be used to trace other galaxy properties, even into the epoch of reionization \citep{cheng2021}.

Most techniques to measuring these target emission lines involve measuring spectra of individually resolved galaxies, making them subject to selection effects including confusion and instrument noise. Confusion occurs when undetected galaxies provide noise to the signal, particularly affecting telescopes with large aperture sizes or low frequency resolution. Instrument noise prevents astronomers from detecting galaxies much fainter than the levels of signal fluctuations in the instrument without unlimited integration time per pixel. Furthermore, time-intensive spectroscopy limits many surveys to smaller volumes, therefore undersampling the rarer, brighter objects. As a result, direct-detection surveys only detect a subset of the total galaxy population given a finite survey time. More nefariously, this may lead to biases in the inference of properties of galaxy evolution. This includes, for example, a bias of directly-measurable bright high-redshift galaxies ($z\gtrsim 2$) toward toward those hosting active galactic nuclei.

Intensity mapping may complement these studies by performing a measurement of the mean line intensity at a given redshift. Here, the large-scale clustering power spectrum is proportional to the first moment of the luminosity function $\langle L\rangle$, whereas the shot noise power spectrum is proportional to the second moment $\langle L^2\rangle$. We stress that both of these statistics include the entire galaxy population, including the faint-end not detected by direct-detection surveys. These first- and second-moment statistics complement direct-detection surveys by sampling the entire population, and featuring unique parameter degeneracies in the line luminosity function.

A number of challenges remain in our ability to construct intensity maps. Light the solar system and galaxy provide a bright foreground signal, which may be many orders of magnitude larger than the target intensity. Fortunately, however, most of these foreground signals vary smoothly with frequency, while the target intensity varies sharply due to matter fluctuations along the line of sight. Given sufficient stability of the instrument passband response and knowledge of the telescope beam, the smooth modes can be effectively removed from the map, negating nearly the entire foreground population. Another challenge is contamination from interloping lines, though this may be more straightforwardly addressed through appropriate masking, modeling, or cross-correlation studies.

Line intensity mapping is a young, but rapidly-growing field. The HI intensity has been detected through cross-correlations between radio telescopes and existing galaxy redshift surveys \citep{chang2010,masui2013,switzer2013,anderson2018,wolz2021}. Similar studies were then performed with other lines, and preliminary detections have been made of [CII] \citep{pullen2018,yang2019}, CO \citep{keating2020,keenan2021}, and Ly$\alpha$ \citep{croft2018}. These first-generation detections have inspired confidence in our ability to overcome challenging systematics, while also providing novel information constraining galaxy evolution models.

In this proceeding, we will model [CII] maps obtained by the EXperiment for Cryogenic Large-aperture Intensity Mapping (EXCLAIM) \citep{cataldo2020}. EXCLAIM is a suborbital mission led by NASA-Goddard Space Flight Center, and is both a technological and scientific pathfinder for intensity mapping aiming to follow up tentative [CII] detections by \citet{pullen2018,yang2019} with a definitive measurement. Various other telescopes are being built targeting [CII] at other redshifts, including CCAT-prime \citep{stacey2018}, Concerto \citep{ade2020}, TIM \citep{vieira2019}, and TIME \citep{sun2021}. Future space telescopes, including the proposed Galaxy Evolution Probe \citep{glenn2018} and Origins Space Telescope \citep{leisawitz2021} may also utilize the intensity mapping technique to map [CII] and various other lines.

We begin by describing the line intensity overdensity, a biased tracer of the matter overdensity, and its calculation based on models of line luminosity functions. We then describe the power spectrum, the primary statistic for LIM, as well as other probes of large-scale structure. Finally, we describe next steps for study, including inferring cosmology and galaxy evolution properties, the effects of foregrounds, and cross-correlation methods.

\textbf{Tracing the Matter Overdensity}

\begin{figure}
    \centering
    \includegraphics[width=\textwidth]{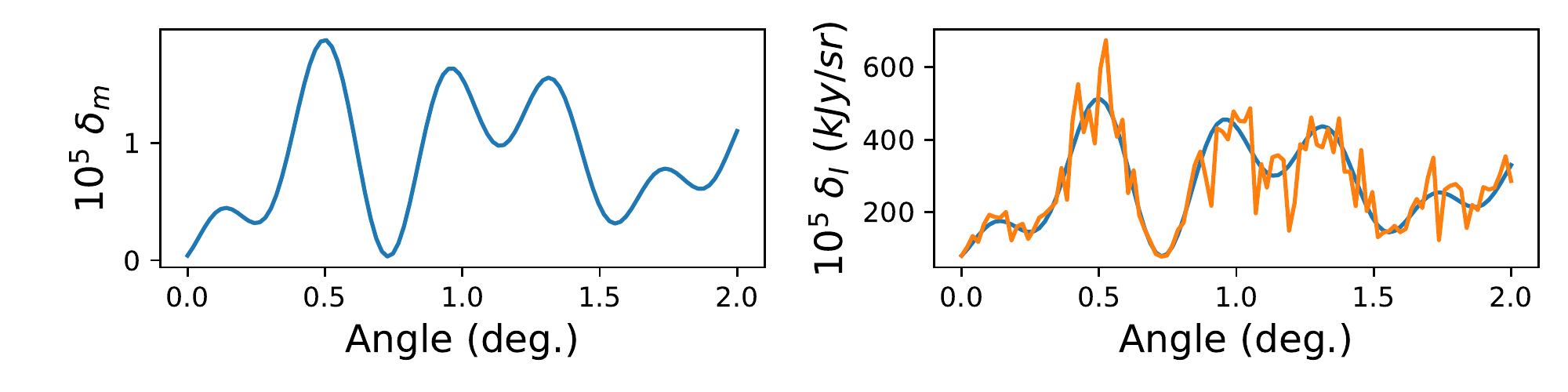}
    \caption{Demonstration of the 1-dimensional CII overdensity at $z=3$. (Left) matter power spectrum; (right) CII intensity calculated by Equation \ref{eqn:delta_I} with (blue) and without (orange) noise. We construct the matter overdensity through the superposition of 10 randomly-selected modes with random phase, with magnitudes determined by the power spectrum. We nominally take the noise amplitude to be $30\%$ of the bias times the intensity.}
    \label{fig:overdensity}
\end{figure}

The primary observable in an intensity map is the intensity field $\delta_I$, which traces large-scale matter fluctuations $\delta_m$ as
\begin{equation}
    \delta_I(\textbf{r},z) = \bar{I}(z) b(z) \delta_m(\textbf{r},z) + N(z),
    \label{eqn:delta_I}
\end{equation}
where $\textbf{r}$ represents spatial location and $b$ is the linear clustering bias of the  observed galaxies relative to the underlying matter overdensity $\delta_m$. $\bar{I}$ is the intensity at mean density, representing the cumulative emission from all galaxies in the survey region. $N$ describes map noise, which we assume is Gaussian and random at a given frequency/redshift.

The intensity $I$ is related to the first moment of the galaxy luminosity function, and its dependence on $z$ probes galaxy evolution. The specific intensity, which we will simply refer to as the intensity, of a single galaxy is given by \citep{carilli2013}
\begin{equation}
    I_{1~\rm gal} = \frac{L}{4\pi d_L^2} \frac{1}{\delta \nu \Omega_{\rm beam}},
\end{equation}
where $L /(4\pi d_L^2 \delta \nu)$ provides a specific flux (units of $W~m^{-2}~Hz^{-1}$), and $1/\Omega_{\rm beam}$ converts from flux to intensity. Here, $d_L$ is the luminosity distance. We define all the terms except $L$ by
\begin{equation}
    \frac{dI}{dL} \equiv \frac{1}{4\pi d_L^2 \delta \nu \Omega_{\rm beam}}.
\end{equation}

An intensity map provides an integral over all galaxies in a given region, resulting in
\begin{eqnarray}
    \bar{I} &=& \int \frac{L}{4 \pi d_L^2 \Omega_{\rm beam}} \frac{\partial V_{\rm co}}{\partial \nu} d n \nonumber \\
    &=& \int \frac{L}{4 \pi d_L^2 \Omega_{\rm beam}} \frac{\partial V_{\rm co}}{\partial z} \frac{\partial z}{\partial \nu} d n \nonumber \\
    &=& \frac{\lambda_{\rm rest}}{4 \pi H(z)} \int L d n,
\end{eqnarray}
where $\partial V_{\rm co} / \partial z = d_A^2 (1+z)^2 \Omega_{\rm beam} c / H(z)$ \citep{hogg1999} and $\partial z / \partial \nu = (1+z)^2 \lambda_{\rm rest} / c$. Here, $dn$ is the differential number of line-emitting galaxies per comoving volume $V_{\rm co}$, $\lambda_{\rm rest}$ is the rest-frame wavelength of the target line, and $d_A=d_L/(1+z)^2$ the comoving angular diameter distance.

The number density of galaxies can be calculated in a number of ways, including
\begin{eqnarray}
    L dn &=& L \frac{dn}{dL}(L) dL ~~~{(\rm Luminosity~ approach)} \nonumber \\
    &=& L(M) \frac{dn}{dM}(M) dM ~~~{(\rm Halo~ mass~ approach)} \nonumber \\
    &=& L(L_{\rm IR}) \frac{dn}{dL_{\rm IR}}(L_{\rm IR}) dL_{\rm IR} ~~~{(\rm IR~ luminosity~ approach)}. \nonumber
    \label{eqn:model_approaches}
\end{eqnarray}

In the luminosity approach, the primary model parameter is the conditional luminosity function $dn/dL$, describing the differential number density of galaxies per luminosity, in units of ${\rm Mpc}^{-3} {\rm L}_\odot^{-1}$. Multiplying by a differential luminosity (in practice, a luminosity bin), we obtain the number density of galaxies at the given luminosity.

In the halo mass approach, we relate galaxy luminosities to an underlying dark matter halo model \citep{cooray2002}. Here, we assume a one-to-one relationship (potentially including scatter) between halo mass $M$ and the luminosity associated with the halo $L(M)$. The differential number density of halos per mass $dn/dM$ is known as the \textit{halo mass function}, and is well-constrained by N-body simulations and analytical calculations \citep{sheth2002}. The halo mass approach is particularly advantageous in its relation to cosmology, where $dn/dM$ is closely related to linear structure formation and $L(M)$ is well-connected to the cosmic star and galaxy formation history, as well as the connection between dark and luminous matter.

In the infrared luminosity approach, the line luminosity is related to the bolometric infrared luminosity $L_{\rm IR}$, typically defined as the integral of all galaxy spectral energy distributions between the rest frequencies $8 - 1000 \,\rm GHz$. Here, the infrared luminosity function $dn/dL_{\rm IR}$ describes the differential number density of galaxies per infrared luminosities and tends to be better-constrained than line luminosity functions. A relation between line luminosity and infrared luminosity is needed, and is typically modeled by observation by a power law \citep{spinoglio2012}. The infrared luminosity approach is advantageous in deriving bulk galaxy properties without direct considerations of the underlying dark matter halo distribution. For example, the cosmic star formation rate is thought to be related to the bolometric infrared luminosity function through a constant of proportionality known as a Kennicutt-Schmidt Relation \citep{kennicutt1998}.

For the rest of this section we will apply the halo mass function approach \citep{cooray2002}. Here,

\begin{equation}
    \bar{I}(z) = \frac{\lambda_{\rm rest}}{4 \pi H(z) } \int_{M_{\rm min}}^\infty L(M,z) \frac{dn}{dM}(M,z) dM,
    \label{eqn:I_M}
\end{equation}
where we have re-instated the $z$-dependence in all terms.

The clustering bias $b$ between the intensity field and the underlying matter density field provides another observable that may be used to infer the mass-luminosity function. We calculate $b$ as a weighted mean over the mass-luminosity function,
\begin{equation}
    b(z) = \frac{\int_{M_{\rm min}} b_h(M,z) L(M,z) dn/dM(M,z) dM}{\int_{M_{\rm min}} L(M,z) dn/dM(M,z) dM}
    \label{eqn:b_M} 
\end{equation}
where $b_h(M,z)$ is the bias of a given halo against the underlying matter power spectrum \citep{sheth2002,tinker2010}.



 While the mean intensity $\bar{I}$ is proportional to the \textit{first moment} of the luminosity function, it is also useful to define the \textit{second moment} of the luminosity function, which we denote by $L_2$, as
 \begin{equation}
     L_2(z) = \int_{M_{\rm min}}^\infty L^2(M,z) \frac{dn}{dM}(M,z) dM,
     \label{eqn:I_2}
 \end{equation}
 given in units of $L_\odot^2 {\rm Mpc}^{-3}$.

\textbf{Power Spectrum Statistics}

Typically, the intensity field $\delta_I$ in Equation \ref{eqn:delta_I} is described through its Fourier Transform,
\begin{equation}
    \tilde{\delta}_I(\textbf{k},z) = \frac{1}{(2\pi)^{3/2}} \int d^3 \textbf{r} \delta_I(\textbf{r},z) e^{i \textbf{k}\cdot\textbf{r}},
\end{equation}
where $\textbf{k}$ is the wavenumber of the spatial fluctuations described by coordinates $\textbf{r}$.

Statistically, the matter fluctuations $\tilde{\delta}_I$ are described by the intensity power spectrum $P_I(\textbf{k},z)$, given by
\begin{equation}
    \langle \delta_I(\textbf{k},z) \delta_I(\textbf{k}',z) \rangle = (2\pi)^3 P_I(\textbf{k},z)  \delta(\textbf{k}-\textbf{k}'),
    \label{eqn:power_spectrum_definition}
\end{equation}
where $\langle \rangle$ denotes an expectation value. Here, $\delta$ denotes the Dirac delta function.

The intensity power spectrum features three primary contributions: a 2-halo (clustering) term $P_I^{\rm clust}$, a 1-halo term $P_I^{\rm 1h}$, and a shot noise term $P_I^{\rm shot}$, i.e.
\begin{equation}
    P_I(\textbf{k},z) = P_I^{\rm clust}(\textbf{k},z) + P_I^{\rm shot}(z)
    \label{eqn:power_spectrum}
\end{equation}

The clustering term describes linear fluctuations tracing the matter power spectrum, on scales larger than individual dark matter halos. Here,
\begin{figure}
    \centering
    \includegraphics[width=.5\textwidth]{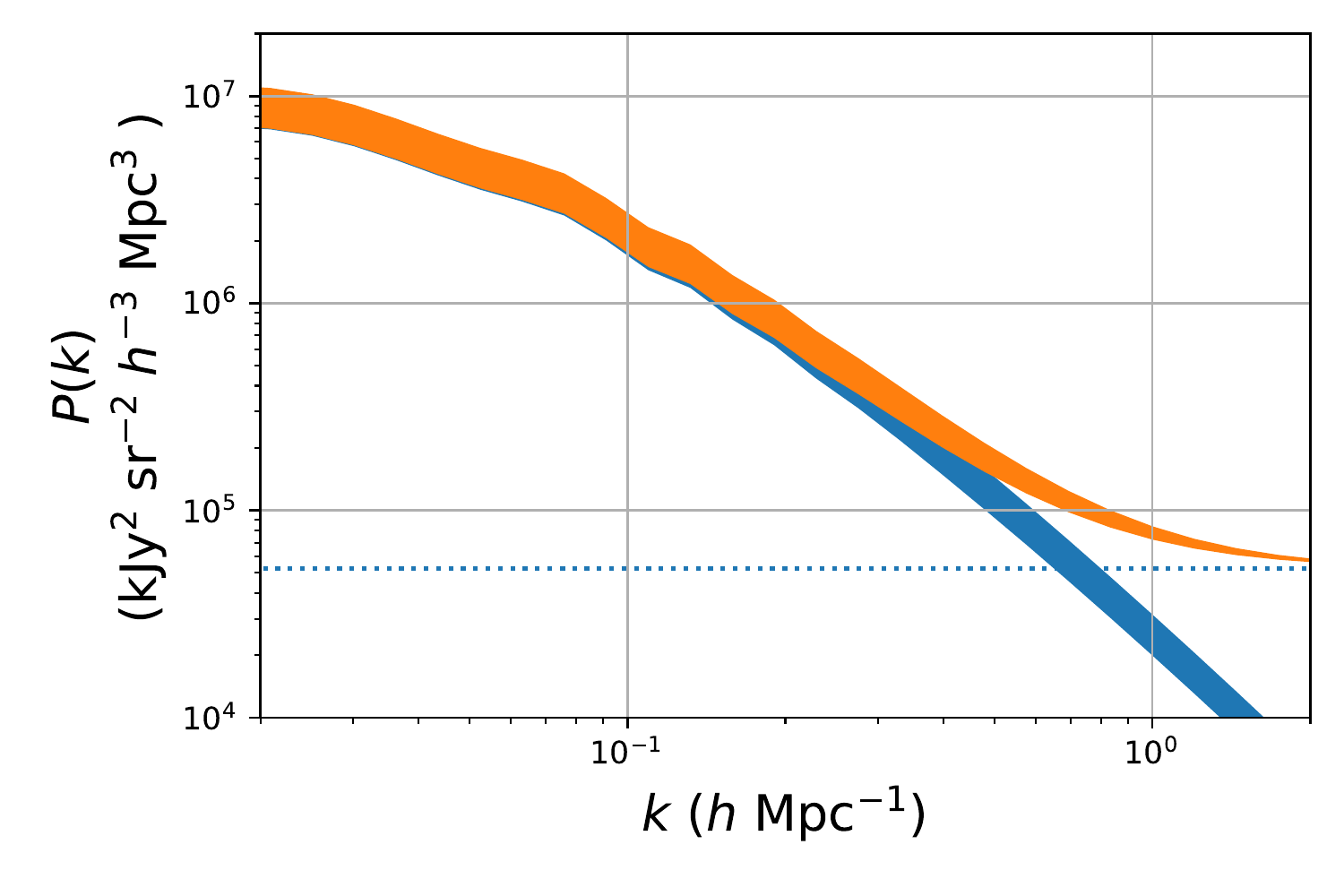}
    \caption{CII power spectrum at redshift $z=3$. Here, the (upper) orange curve shows the total power spectrum, the blue curve shows the clustering power spectrum, and the dotted line shows the shot power.}
    \label{fig:power_spectra}
\end{figure}

\begin{equation}
    P_I^{\rm clust}(\textbf{k},z) = \bar{I}^2(z) b^2(z) F_{\rm RSD}^2(\textbf{k},z) P_m(k,z),
\end{equation}
where $P_m$ is the power spectrum of the matter overdensity $\delta_m$, defined analogously to Equation \ref{eqn:power_spectrum_definition}.

\begin{equation}
    F_{\rm RSD}(\textbf{k},z) = \left ( 1 + \frac{f}{b}\mu^2 \right ) \exp \left [ -\frac{1}{2}\mu^2 k^2 \sigma_k^2 \right ],
\end{equation}
where $\sigma_k=$, $\mu \equiv k_{||}/k$. Here, the first term represents the \textit{Kaiser Effect} \citep{kaiser1987}, representing redshift-space distortions due to the peculiar velocity of galaxy-hosting halos, i.e. their velocities relative to the Hubble flow. The exponential term describes \textit{Fingers of God} distortions due to velocity dispersion of line-emitting galaxies within host halos.

The shot power term describes Poisson errors due to the fact that line-emitting galaxies follow discrete statistics, given by (see Appendix A of \citep{breysse2019} for a derivation)
\begin{equation}
    P_I^{\rm shot}(z) = \left ( \frac{\lambda_{\rm rest}}{4\pi H(z)} \right )^2 I_2(z),
\end{equation}
with $L_2$ given by Equation \ref{eqn:I_2}.

Figure \ref{fig:power_spectra} shows the power spectrum calculated by the methods described here. We model $dn/dM$ through COLOSSUS software \citep{diemer2018}, and use the [CII] mass-luminosity function from \citet{padmanabhan2019}, resulting in $\bar{I}(z=3) = 15.7\,\rm kJy~\rm sr^{-1}$ and $b(z=3) = 3.48$. The orange (upper) shaded curve shows the total power spectrum, including both clustering and shot power spectra, the dark blue (lower) shaded curve describes the clustering power spectrum, and the flat dotted curve shows the shot power. The shaded regions describe the range of $\mu$ values, where the lower limits describe the case where we only obtain modes perpendicular to the line of sight, i.e. $\mu=0$, whereas the upper limit describe modes parallel to the line of sight, where $\mu=1$.


\textbf{Next Steps: Noise, Foregrounds and Cross-Correlations}

We note that Equations \ref{eqn:delta_I} and \ref{eqn:power_spectrum} are simplifications compared to realistic signals. Foregrounds present a principal challenge; here, they add additive terms to Equation \ref{eqn:delta_I} and \ref{eqn:power_spectrum_definition} representing the foreground intensity $F$ and power spectrum $P_F$, respectively. Zodiacal light tends to dominate near-infrared observations, milky way Cirrus emission and the cosmic microwave background dominate the mid-to-far infrared relevant to [CII], while galactic synchrotron radiation and point sources dominate the radio. In all of these cases, the foregrounds are expected to vary much more smoothly than the target LIM signal, while spatial fluctuations are still strong on all scales. Therefore, in $k$-space, the foregrounds are expected to dominate lower-$k_{||}$ modes. Given a sufficiently stable passband, the foregrounds can be filtered by removing the brightest $k_{||}$ modes, without a strong loss of the target power spectrum. Methods for single-dish observations can be found in \citet{switzer2013,anderson2018}.

Instrument noise presents an additional term that adds variance to the intensity map. While this calculation depends on the specifics of the target line and instrument and is beyond the scope of the current proceeding, we refer the reader to \citet{bernal2019,oxholm2021} for equations and background. These references also describe the formalism for cosmological forecasts with LIM experiments, relying on the Fisher matrix \citep{tegmark1997}.

Foregrounds and instrument noise present additive biases to the intensity mapping signal $\delta_I$ which can be removed through cross-correlation with a second large-scale structure survey. Here, the second survey must trace the same matter density field $\delta_m$, i.e. it must occupy the same volume as the intensity mapping survey. For example, a galaxy redshift can be described through a galaxy density field $\delta_g(\textbf{r},z)=b_g(z) \delta_m(\textbf{r},z)$. Here, cross-correlation with the intensity mapping field yields
\begin{eqnarray}
    \langle \delta_I(\textbf{r},z) \delta_g(\textbf{r}',z) \rangle &=& \langle \left ( \bar{I}(z) b(z) \delta_m(\textbf{r},z) + N(z) + F(\textbf{r},\nu_z) \right ) b_g \delta_m(\textbf{r}',z) \rangle \nonumber \\
    &=& \bar{I}(z) b(z) b_g(z) \langle \delta_m(\textbf{r},z) \delta_m(\textbf{r}',z) \rangle,
\end{eqnarray}
where the $N$ and $F$ terms are canceled because they do not correlate with the density field. We note, however, that the other terms are still present in the variance of the cross-power spectrum. Other sources of noise in an intensity map, including contamination from interloping lines at other redshifts \citep{lidz2016,gong2020}, may also be mitigated through cross-correlation. Further details on intensity mapping cross-correlations, including their power spectra and their ability to cancel cosmic variance, are also described in \citet{oxholm2021}.

\vspace{1cm}
\textbf{Conclusion}

We have described the formalism necessary to model intensity mapping power spectra. We modeled a power spectrum plot for the [CII] intensity map at redshift $z=3$, representing a target emission for EXCLAIM, a first-generation dedicated intensity mapping instrument. Furthermore, we described crucial next steps of implementing instrument noise and foregrounds, the principal sources of variance in LIM measurements. Once these systematics are overcome, LIM power spectra may be used to dramatically increase our understanding of galaxy evolution and cosmology from present day and into the dark ages.



\def\bibfont{\footnotesize}

\vspace{-0.7cm}
\bibliography{refs}

\end{document}